\documentclass{webofc}
\usepackage[varg]{txfonts}   
\usepackage[colorlinks=true,linkcolor=black]{hyperref}
\newcommand{\nn}{\nonumber}
\newcommand{\vp}{{\mathbf{p}}}
\newcommand{\lm}{\lambda}

\newcommand{\ra}{\rangle}
\newcommand{\cl}{{\cal L}}

\begin{document}
\title{Unitarizing infinite-range forces: Graviton-graviton scattering, the graviball, and Coulomb scattering}
%
%

\author{\firstname{Jos\'e Antonio} \lastname{Oller}\inst{1}\fnsep\thanks{\email{oller@um.es}} 
}

\institute{Departamento de F\'{\i}sica, Universidad de Murcia, 30071 Murcia, Spain}

\abstract{%
 We study graviton-graviton scattering in partial-wave amplitudes after unitarizing their Born terms. In order to apply $S$-matrix techniques, based on unitarity and analyticity, we introduce an $S$-matrix associated to this resummation that is free of infrared divergences. This is achieved by removing the diverging phase factor calculated by Weinberg that multiplies the $S$ matrix, and that stems from the
virtual infrared gravitons. A scalar graviton-graviton resonance with vacuum quantum numbers is obtained as a pole in the nonperturbative $S$-wave amplitude, which is called the graviball. Its resonant effects along the physical real $s$-axis may peak at values substantially lower than the UV cutoff squared of the theory, similarly to the $\sigma$ resonance in QCD. These techniques are also applied to study nonrelativistic Coulomb scattering up to next-to-leading order in the unitarization program. A comparison with the exact known solution is very illuminating.
}
\maketitle
\section{Introduction}
\label{intro}
At energies much smaller than the Planck Mass $M_P^2=G^{-1}$, with $G$ the Newton constant, one can write down a generic action for gravitational interactions organized as a tower of operators with an increasing number of derivatives \cite{Donoghue:1993eb,Donoghue:2017pgk}
\begin{align}
  \label{221013.1}
    S_{\rm grav}&=\int d^4 x\sqrt{-g}\left\{\frac{2}{\kappa^2}R+c_1 R^2+c_2 R_{\mu\nu}R^{\mu\nu}+c_3 R_{\mu\nu\alpha\beta}R^{\mu\nu\alpha\beta}
    +\ldots\right\}~,
\end{align}
where $\kappa^2=32\pi G$, $R_{\mu\nu\alpha\beta}$ is the curvature tensor, $R_{\mu\nu}$ the Ricci tensor and $R$ the curvature. A derivative counts as ${\cal O}(p)$, with $p$ a typical external momentum so that each $R,\, R_{\mu\nu},\, R_{\mu\nu\alpha\beta} \sim p^2$. In addition, every graviton takes a factor $G^{1/2}$ because of the typical splitting of the total metric $g_{\mu\nu}=\eta_{\mu\nu}+\kappa h_{\mu\nu}$, with $\eta_{\mu\nu}$ the Minkowski metric and $h_{\mu\nu}$ the perturbation linear in the graviton field. In the case of pure gravity the counterterms $c_i=0$, $i=1,2,3$, and corrections are ${\cal O}(p^3)$ and higher \cite{vanNieuwenhuizen:1976vb}. Then, for energies $E\ll \Lambda$ with $\Lambda\sim G^{-1/2}\approx 10^{19}$~GeV, gravity can be seen as a low-energy Effective Field Theory (EFT). Here we have identified the cutoff of the theory with the so-called unitary cutoff $\Lambda_U$ of the gravity low-energy EFT \cite{Aydemir:2012nz,Blas:2020och,Blas:2020dyg}.

\begin{figure}[h]
  \centering
  \sidecaption
  \includegraphics[width=.6\textwidth,angle=0]{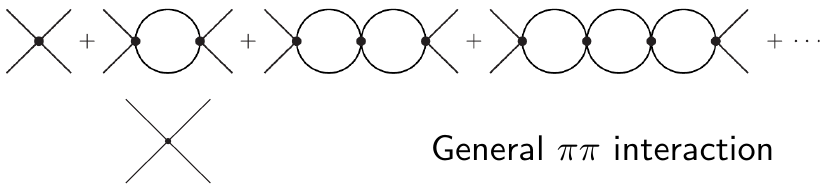}
  \caption{Unitarity loop diagrams resummed by unitarization techniques}
\label{fig-1}       
\end{figure}

For QCD one also has a low-energy EFT in order to describe the interactions between pions, which is called Chiral Perturbation Theory (ChPT) \cite{Weinberg:1978kz,Gasser:1983yg}. In the $SU(2)$ chiral limit, with zero $u$ and $d$ quark masses, the pions are Goldstone bosons and their interactions can also be organized in terms of operators with an increasing number of derivatives, $i\partial_\mu\sim p_\mu$. For this EFT the unitary cutoff  $\Lambda_U=4\pi f_\pi\simeq 1.2$~GeV, with $f_\pi$ the weak pion decay constant, while the cutoff associated with the higher states in QCD is $\Lambda=M_\rho\simeq 0.8$~GeV, with $M_\rho$ the mass of the  $\rho(770)$~\cite{ParticleDataGroup:2022pth}. Below such energies, even more when taking the Mandelstam variable $s$, 
one finds the resonance $f_0(500)$ aka $\sigma$ \cite{ParticleDataGroup:2022pth}, such that $m_\sigma^2/(4\pi f_\pi)^2\approx 0.14$. The smallness of this number makes possible to afford the study of this resonance by applying ChPT to a properly chosen interaction kernel by implementing so-called unitarization techniques \cite{Truong:1988zp,Dobado:1989qm,Oller:1997ti,Oller:1998zr,Oller:1998hw,Nieves:1999bx,Oller:1999ag,Albaladejo:2010tj}, see Ref.~\cite{Oller:2020guq} for a recent review. Schematically the unitarization techniques allow to resum the set of diagrams depicted in Fig.~\ref{fig-1}

In this way, Ref.~\cite{Albaladejo:2012te} by unitarizing the next-to-leading order (NLO) $\pi\pi$ scattering amplitude in partial-wave amplitudes (PWAs) obtained the pole position of the $\sigma$ resonance at $\sqrt{s_\sigma}=458\pm 14-i\,(261\pm 17)$~MeV. This pole position is compatible at the level of one sigma with  $\sqrt{s_\sigma}=457^{+14}_{-13}-i\,(297^{+11}_{-7})$~MeV, the value obtained by solving Roy-like equations  \cite{Garcia-Martin:2011iqs} (the so-called GKPY equations). 
It is also clear from the fact that $|s_\sigma/(4\pi f_\pi)^2|\approx 0.22\ll 1$ 
that the $\sigma$ meson should be actually considered as a low-energy degree of freedom in the EFT and should be accounted for. This point is certainly achieved by employing unitarization techniques, so that this resonance emerges dynamically as a result of the pion interactions.
As explained in Ref.~\cite{Oller:1998zr}, there is a parametrical enhancement affecting the leading-order (LO) isoscalar-scalar $\pi\pi$ PWA  which causes the emergence of the $\sigma$ resonance. 

The $\sigma$ resonance was clearly observed experimentally in the decay $D^+\to\pi^-\pi^+\pi^+$ \cite{E791:2000vek} whose histogram, affected by the strong $\pi\pi$ final-state interactions, was first studied in agreement with the known strong interaction data in Ref.~\cite{Oller:2004xm}. High statistics data on $J/\psi\to \omega\pi^+\pi^-$ also confirmed the presence of the $\sigma$ resonance \cite{BES:2004mws} (see Ref.~\cite{Meissner:2000bc} for a theoretical study of this type of decays). Other important physical magnitudes strongly affected by this low-lying resonance are the vacuum structure, as characterized by the excitations of the quark condensate through the scalar form factor of the pion,  the $\pi$-nucleon $\sigma$ term, the large corrections to the current-algebra prediction of the $J^{PC}=0^{++}$ $\pi\pi$ scattering lengths,  the isoscalar scalar $\pi\pi$ phase shifts, the already commented two-pion event distributions from heavy-meson decays, etc.

Coming back to gravity, let us note the analogies between the low-energy (relative to their respective unitary cutoffs $\Lambda_U$) $\pi\pi$ and graviton-graviton  interactions. Both stems from derivative couplings and its treatment is within the EFT paradigm, cf. Eq.~\eqref{221013.1}. Then, a natural question that emerges is whether there is an analogous  relatively low-lying resonance in graviton-graviton scattering, since one also expects there to have attractive interactions. This question was addressed in Refs.~\cite{Blas:2020och,Blas:2020dyg} in the affirmative sense, and the resulting resonance was called the graviball and its pole position $s_P$. 
Here we are going to review on this finding, the method developed for unitarizing infinite-range interactions, and  the illustration of the latter when applied to nonrelativistic Coulomb interactions, whose solution is exactly known \cite{Oller:2022tmo}.

\section{Infrared-safe PWAs}
\label{sec.221015.1}


If one considers the $t$-channel exchange of a massless particle one expects the appearance of infrared divergences when calculating the PWAs because the angular integration diverges.
 To illustrate this divergence it is enough to show that
\begin{align}
  \label{221015.1}
    &-\int_{-1}^{+1}\frac{dcos\theta}{t}=\frac{1}{2p^2}\left[\log 2-\lim_{\theta\to 0}\log(1-\cos\theta)\right]\to\infty\\
    t&\equiv (\vp-\vp')^2=-2\vp^2(1-\cos\theta)~,~p\equiv |\vp|~.\nn
\end{align}
This divergence is due to the exchange of a {\it virtual soft} photon or graviton ($t\to 0$) 
between two external on-shell lines \cite{Weinberg:1965nx}.

  Dalitz studied  in Ref.~\cite{Dalitz:1951} the potential scattering of an electron with a Yukawa potential $V(r)={e_1 e_2e^{-\mu r}}/{4\pi r}$ in the limit $\mu\to 0$ both in the nonrelativistic and relativistic cases, the former up to ${\cal O}(e^3)$ and the latter up to ${\cal O}(e^2)$. He then conjectured that the infrared divergences summed up giving rise to a  phase factor. 
 The raise of this phase was demonstrated by Weinberg in Ref.~\cite{Weinberg:1965nx} for QED and gravity. In the limit of potential scattering the full extent of virtual infrared photons for any process $\alpha\to\beta$ is to give rise to  a phase factor in the  $S$-matrix given by
\begin{align}
  \label{221015.3}
    \frac{S_{\beta \alpha}}{S_{\beta\alpha}^0({\cal L})}=\exp\left\{\frac{1}{2}\int_\mu^{\cal L} \Im A(q)\right\}~,
\end{align}
where $S_{\beta\alpha}^0({\cal L})$ has no infrared divergences and  $A(q)$ is given in Ref.~\cite{Weinberg:1965nx}. The scale ${\cal L}$ is introduced in this reference to separate between soft and hard physics. The integral in Eq.~\eqref{221015.3} is worked out in Ref.~\cite{Weinberg:1965nx}, such that each different pair of particles $n$ and $m$ in the initial or final state contributes to the $S$ matrix with a diverging phase factor, which has then to be summed over all these pairs. The neat result is
\begin{align}
  \label{221015.3}
\frac{S_{\beta \alpha}}{S_{\beta\alpha}^0({\cal L})}&=S_c=\exp\left\{\frac{i}{2\pi}\sum_{\rm pairs}\frac{e_ne_m}{\beta_{nm}}\log\frac{\mu}{{\cal L}}\right\}~, 
\end{align}
 with $\beta_{nm}=\frac{\sqrt{(p_n\cdot p_m)^2-(m_n m_m)^2}}{p_n\cdot p_m}$~,~the invariant relative velocity between the particles $n$ and $m$ of momenta $p_n$ and $p_m$, respectively. From this equation it is clear that the scale ${\cal L}$ must be proportional to $2p$ since the resummed exchange of soft photons of mass $\mu$ (which acts as an infrared regulator) implies a left-hand cut (LC) starting at $4p^2=-\mu^2$. We then write
 \begin{align}
  \label{220709.24}
   {\cal L}=\frac{2p}{a}~,
 \end{align}
 with the constant $a$ being necessarily independent of $p$, and only enters logarithmically with $\ln a={\cal O}(1)$ \cite{Blas:2020och,Blas:2020dyg}.   For graviton-graviton scattering the analogous to Eq.~\eqref{221015.3} is
\begin{align}
  \label{221016.1}
S_c& \equiv \exp\left\{-i\sum_{\rm pairs}\frac{G m_n m_m (1+\beta_{nm}^2) }{\beta_{nm}[1-\beta_{nm}^2]^{1/2}}\log\frac{\mu}{{\cal L}}\right\}~.
\end{align}

 The next step is to remove the infrared diverging phase factor $S_c$ giving rise to a new $S$-matrix \cite{Kulish:1970ut}. Notice that $S_c$ has no angular dependence and it commutes with a partial-wave expansion of the $S$ matrix, which from now on we assume.  
 If we call the original $S$ matrix affected by the infrared divergences by $\bar{S}_J$, with $J$ a generic subscript indicating the PWA, the new one is denoted by $S_J$ and they are related as in Eq.~\eqref{221015.3}, making the substitution $S_{\beta\alpha}\to \bar{S}_J$ and $S_{\beta\alpha}^0\to S_J$. Namely, for graviton-graviton and Coulomb scattering one has the relations, respectively, 
 \begin{align}
  \label{221016.2}
 S_{J} =S_c^{-1} \bar{S}_J&=\exp\left\{2i Gs\log\frac{\mu}{{\cal L}}\right\}\bar{S}_J~,\\
S_{J} =S_c^{-1} \bar{S}_J&=  \exp\left\{ \frac{2i\alpha}{\beta}\log\frac{\mu}{{\cal L}}\right\}\bar{S}_J~,\nn
 \end{align}
 with $\alpha=e^2/4\pi$, the fine-structure constant. Associated with an infrared-finite $S_J$ we have the PWA $T_J$, related by  $S_J=1+\frac{i\pi}{4}T_J$~, and $S_J=1+\frac{im p}{\pi}T_J$~, 
  for graviton-graviton and nonrelativistic Coulomb scattering, respectively. 

 \section{Graviton-graviton scattering in PWAs and the graviball}
 \label{sec.221016.1}


 The state of a free graviton is determined by its momentum $\vp$ and helicity $\lambda$.
 The Born terms $F_{\lambda_3\lambda_4,\lambda_1\lambda_2}$ for the scattering of two gravitons $|\vp_1,\lambda_1\ra |\vp_2,\lambda_2\ra \rightarrow  |\vp_3,\lambda_3\ra |\vp_4,\lambda_4\ra $,
 are taken from Ref.~\cite{grisaru.170513.1}. The non-zero Born-term amplitudes can be found there or in Ref.~\cite{Blas:2020dyg}.  
In order to end with infrared-free PWAs we have to take into account the factor $S_c^{-1}$ in Eq.~\eqref{221016.2} up to ${\cal O}(G)$. 
This implies,
\begin{align}
  \label{221016.5}
  S^{(J)}_{\lambda_3\lambda_4,\lambda_1\lambda_2}&=\left[1+\frac{i\pi 2^{|\lambda|/4}}{4}F^{(J)}_{\lambda_3\lambda_4,\lambda_1\lambda_2}\right]
  \left[1-2iGs\log\frac{{\cal L}}{\mu}\right]
  +{\cal O}(G^2)
  \equiv 1+\frac{i\pi}{8}V^{(J)}_{\lambda_3\lambda_4,\lambda_1\lambda_2}~,
\end{align}
where $|\lambda|=|\lambda_2-\lambda_1|$. 
For instance, for $F_{22,22}^{(0)}$ we have that
\begin{align}
  \label{221016.6}
F^{(0)}_{22,22}(s)&=-\frac{\kappa^2 s^2}{16\pi^2}\int_{-1}^{+1} 
\frac{d\!\cos\theta}{t-\mu^2}
\to  \frac{8Gs}{\pi}\log\frac{2p}{\mu}~,\\
V^{(0)}_{22,22}(s)&=\frac{8G s}{\pi}\log a~.\nn
\end{align}

Once we have at our disposal the infrared-safe PWAs we proceed to its unitarization by employing standard techniques in hadron physics. In this way, for calculating the unitarized PWA $T^{(J)}_{\lm_3\lm_4,\lm_1\lm_2}(s)$ the unitary or right-hand cut can be resummed by employing the general expression
\begin{align}
  \label{221016.6}
2^{|\lm|/4}T^{(J)}_{\lm_3\lm_4,\lm_1\lm_2}(s)=\left[\frac{1}{2^{|\lm|/4}V_{\lm_3\lm_4,\lm_1\lm_2}^{(J)}}+\frac{1}{8}\ln\frac{-s}{\Lambda^2}\right]^{-1}~, 
\end{align}
where $\Lambda$ is the cutoff of the low-energy gravity EFT. This expression arises by performing a dispersion relation of the inverse of the PWA with a circle in infinity deformed to engulf the right-hand cut \cite{Blas:2020dyg}. In the subsequent, we identify $\Lambda$ with the unitary cutoff $\Lambda_U^2=\pi/(G \ln a)$ and look for poles of $T^{(0)}_{22,22}(s)$ in the second RS, 
which implies the following secular equation for the pole positions,
\begin{align}
  \label{221016.8}
     &\frac{1}{x}+\log(-x)-i2\pi=0~,~x=\frac{s_P}{\Lambda^2}~.
\end{align}
It can be easily solved by iteration or numerically,  $x=0.07-i\,0.20\simeq -i2/(3\pi)$. The estimated uncertainty on $x$ from higher order contributions is a 20\% \cite{Blas:2020och,Blas:2020dyg}. The reason is because the main source of expected uncertainty stems from the one-loop or NLO graviton-graviton PWA, which has a size that scales as $|x|=|s_P|/\Lambda_U^2$.

The dependence on the cutoff $\Lambda^2\neq \Lambda_U^2$ can be followed by introducing the parameter $\omega=\Lambda^2/\Lambda_U^2$. In terms of it the generalization of  Eq.~\eqref{221016.8} reads ${1}/{\omega x}+\ln(-x)-i2\pi=0$~. The modulus of its solution roughly scales like $\sim 1/\omega$. 
As a result, when $\omega>1$ (so that $\Lambda>\Lambda_U$), $x$ decreases. However, in absolute terms $s_P=x \Lambda^2$ remains rather constant, decreasing only very slowly  as $\omega$ increases (for instance, for $\omega=10$ its modulus is $0.18~\pi G^{-1}$ instead of $0.22~\pi G^{-1}$ for $\omega=1$).  In turn, for $\omega<1$ ($\Lambda^2<\Lambda_U^2$) then $x$ increases. This is troublesome because it implies that  $|s_P|$ will become comparable  with $\Lambda^2$ as $\omega$ decreases. This would  invalidate our approximate calculation of the interaction kernel $V_{\lm_3\lm_4,\lm_1\lm_2}^{(J)}(s)$ in perturbation theory, which is then used to find the corresponding $T^{(J)}_{\lm_3\lm_4,\lm_1\lm_2}(s)$ by unitarizing the former.

At this point, we recall that Ref.~\cite{Blas:2020dyg} also studied in similar terms the pole position of the $\sigma$ resonance by unitarizing the LO isoscalar scalar $\pi\pi$ PWA in the chiral limit, and the same secular equation as Eq.~\eqref{221016.8} was found.
 It is also instructive to compare its numerical solution  $x_\sigma=0.07-i\,0.20$ (the same one as given above for the graviball),   with the value from the GKPY equations for the actual physical pion mass, $x_\sigma=0.09-i\,0.20$, being remarkably close to each other.

\begin{figure}[h]
\centering
  \sidecaption
\includegraphics[angle=0.0,width=.5\textwidth]{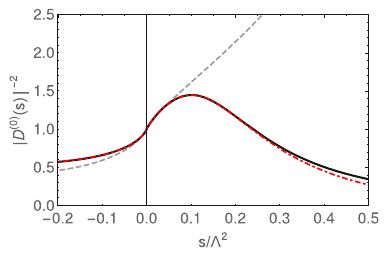}
  \caption{Modulus squared of the Omn\`es function $\Omega^{(0)}(s)$.}
\label{fig.221016.2}       
\end{figure}

 Now, the smallness of the real part of $x_P$, 0.07, versus the modulus of its imaginary part, $0.20$,  implies that the resonance effects of the graviball (as well as those of the $\sigma$) peak in the variable $s$ at much lower values than $\Lambda_U^2$. This can be explicitly shown by considering the Omn\`es function $\Omega^{(0)}(s)=T^{(0)}/V^{(0)}(s)\equiv D^{(0)}(s)^{-1}$, whose modulus squared in shown in Fig.~\ref{fig.221016.2}.  Reference~\cite{Blas:2020dyg}  also argued about the numerical suppression of the phase space of multi-graviton states, which for $s<\Lambda_U^2$ provides a strong suppression of this type of multi-particle states in the unitarity relation.  This is a similar effect to that discussed for $\pi\pi$ scattering in Ref.~\cite{Salas-Bernardez:2020hua}.   Additionally, Ref.~\cite{Blas:2020dyg} considered the impact of higher-order monomials in the expansion of Eq.~\eqref{221013.1}, containing three and four powers of the curvature tensor (and its contraction thereof). The former are indicated by $\{R^3\}$ and the latter by $\{R^4\}$. It was found in Ref.~\cite{vanNieuwenhuizen:1976vb} that the $\{R^3\}$ terms do not contribute to the scattering amplitude $T_{22,22}$, while the $\{R^4\}$ do. 
 These extra terms give rise to a relative change in the pole position of the graviball of around a 3\% \cite{Blas:2020dyg}, in agreement with the expected $x^3=1\%$.

A way to make lighter the graviball is to couple gravitons with light matter fields (of mass much smaller than $G^{-1/2}$). Roughly this will imply to multiply by $N$ the unitary loop function $\ln(-s/\Lambda^2)$ in Eq.~\eqref{221016.6}, so that the secular equation becomes $1/(N x_N)+ \ln(-x_N)-i2\pi=0$, whose approximate solutions scales as $1/N$. Then $s_P=x_N \pi G^{-1}$ decreases as $\sim 1/N$, which makes the graviball lighter and narrower. In some scenarios \cite{Dvali:2014ila,han.200204.1,Arkani-Hamed:2016rle} it is expected that $\Lambda^2$ also decreases as $1/N$ so that in this case $x_N\sim x$ and it does not evolve with $N$. However, at the absolute level, once $x$ is multiplied by $\Lambda^2$, again $s_\sigma\sim 1/N$. Therefore, in theories with large enough number of light degrees of freedom (or with large extra dimensions)   the graviball could affect gravitational phenomena at scales that could be tested in colliders (see Ref.~\cite{Blas:2020dyg} and references thereof).

Reference~\cite{Blas:2020dyg} also studied the presence of the graviball at dimensions $d$ larger than 4, in which the graviton-graviton scattering is infrared safe. The graviball still persists for $d>4$ and, by imposing the smoothest transition from $d\geq 5$ down to $d=4$, one could have a way to estimate the parameter $\ln a$. In this way, it was obtained that $\ln a\approx 1$, which allows to give $s_\sigma$ in absolute terms as $s_\sigma=(0.22-i\,0.63)~G^{-1}$ with an estimated uncertainty of a 20\%. Reference \cite{Guerrieri:2021ivu} also identifies in $d=10$ supergravity a prominent scalar resonance as the lightest one in graviton-graviton scattering. 

\section{Unitarizing Coulomb scattering}
\label{sec.221017.1}

Essentially the same method as outlined in the previous section for graviton-graviton scattering was also applied in Ref.~\cite{Oller:2022tmo} to study the unitarization of  nonrelativistic Coulomb scattering up to NLO. In this case, the phase factor $S_c=\exp(2i\gamma \ln{\cal L}/\mu)$, with $\gamma=m\alpha/p$. From an expansion analogous to Eq.~\eqref{221016.5} but up to NLO, the infrared-free PWAs at LO and NLO are, respectively,  
\begin{align}
  \label{221017.2}
  T_J^{(1)}&=F_J^{(1)}(p)-\frac{e^2}{2p^2}\ln\frac{{\cal L}}{\mu}~,\\
  T_J^{(2)}&=F_J^{(2)}(p)-i F_J^{(1)}(p)\frac{me^2}{2\pi p}\log\frac{\cl}{\mu}
+i\frac{me^4}{8\pi p^3} \left(\log\frac{\cl}{\mu}\right)^2 ~.\nn
\end{align}

We now particularize to the case $J=0$. Then, it is straightforward to end with $F_0^{(1)}(p)=\frac{e^2}{2p^2}\ln\frac{2p}{\mu}$ and $T_0^{(1)}(p)=\frac{e^2}{2p}\ln a$. The technical details for the NLO perturbative amplitude can be found in Ref.~\cite{Oller:2022tmo}, where a straightforward calculation gives that $F_0^{(2)}(p)=i \frac{mp}{2\pi}{F_0^{(1)}}^2$. For the nonrelativistic kinematics the unitarization formula  analogous to Eq.~\eqref{221016.6} up NLO reads
\begin{align}
  \label{221017.3}
  T_J&=\left[\frac{1}{V_0^{(1)}+V_0^{(2)}}-i\frac{mp}{2\pi}\right]^{-1}
=V_0^{(1)}+ V_0^{(2)}+i\frac{mp}{2\pi}{V_0^{(1)}}^2+{\cal O}(\alpha^3)~,
\end{align}
from where $V_0^{(1)}=\frac{e^2}{2p^2}\ln a~$ and $V_0^{(2)}=0$ by matching with $T_0=T_0^{(1)}+T_0^{(2)}+{\cal O}(\alpha^3)$.  The exact $S$ matrix in PWAs is known for Coulomb scattering, and it reads
\begin{align}
  \label{221017.4}
  S_J&=\frac{\Gamma\left(1+J-i\gamma \right)}{\Gamma\left(1+J+i\gamma \right)}~.
  \end{align}
By equating it with $S_J=1+i m p T_J/\pi$ we can also deduce the expression for  $T_J$, and from that the resulting unitarization kernel $V_J$ to all orders
\begin{align}
  \label{221017.5}
        V_J&=\frac{2i\pi}{m_r p}\frac{\Gamma(1+J+i\gamma)-\Gamma(1+J-i\gamma )}{\Gamma(1+J+i\gamma)+\Gamma(1+J-i\gamma)}~.
\end{align}

     \begin{table}
       \centering
         \caption{Pole positions in the complex $p$-plane corresponding to the ground state. 
           The pole position is given with respect to the exact value $im\alpha$.
           \label{tab.221017.1}}
         \begin{tabular}{|l|llllll|}
           \hline
           $n$ & 1 & 3 & 5 & 7 & 9 & 11\\
           \hline
           ${p^{(n)}}/p_{\rm exact}$ & $0.577$ & $0.950$ & $0.998$ & $1.003$ &
           $1.002$ & $1.001$  \\
\hline
         \end{tabular}
       \end{table}

     Working out the expansion in powers of $\gamma$ or $\alpha$ for $V_0$, one has that $V_0=-2\pi \alpha\gamma_E/p^2+{\cal O}(\alpha^3)$, in agreement with our perturbative calculation of $V_0^{(1)}$ and $V_0^{(2)}=0$, fixing then $\ln a=\gamma_E=0.577$. There is another way to fix $\ln a$ by using the known asymptotic behavior of the Coulomb wave functions, which was considered in Ref.~\cite{Blas:2020dyg}. Within the small angle approximation to all orders in  $\alpha$, Ref.~\cite{Bazhanov:1977fa} in QED also obtains the extra phase $-2\gamma(\log \mu+\gamma_E)$ of $S_c$.

\begin{figure}[ht]
       \centering
  \sidecaption
         \includegraphics[width=.5\textwidth,angle=0]{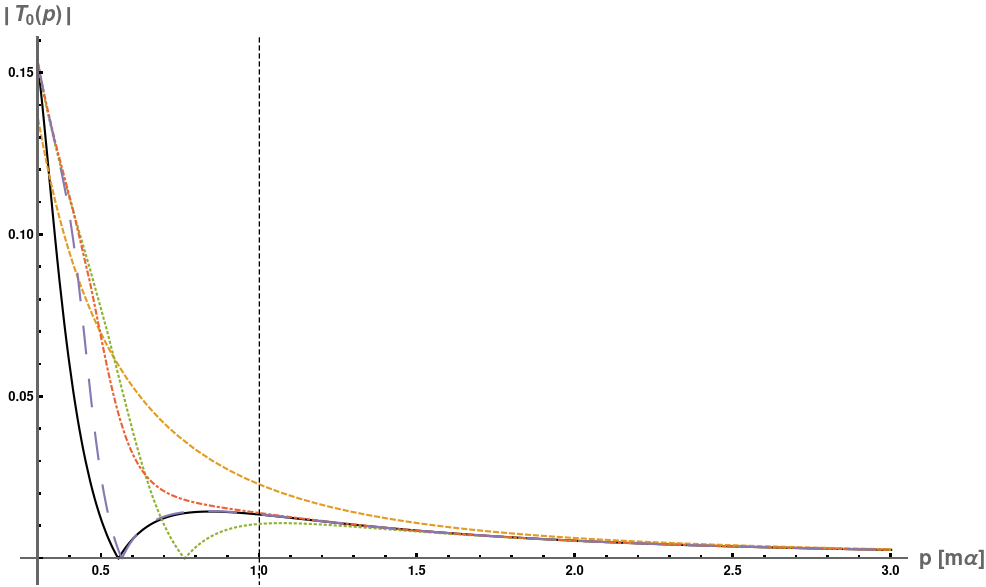}
  \caption{\small The modulus of the Coulomb  $S$-wave PWA given by the exact result (black solid line) is compared with those obtained by applying the unitarization formula up to order ${\cal O}(\alpha^n)$ in the calculation of $V_0=\sum_{i=1}^n V^{(i)}_0(p)$.
    The lines with $n=1,$ 3, 5 and 7 are given by the orange dashed, green dotted, red dash-dotted and  magenta long-dashed lines, respectively. In the figure energy units are taken such that $m\alpha=1$, with the vertical line indicating this value for $p$. \label{fig.220711.1}}
     \end{figure}

     One can go on with the expansion of $V_0$ in odd powers of $\alpha$ (the even powers give zero since it is an odd function in $\alpha$), and work out the pole position of the ground state of the Hydrogen atom by determining the pole position of the resulting unitarized $T_0(p)$ amplitude. The results are shown in Table~\ref{tab.221017.1} for $n=1$ up to  11, and we see a rapid convergence towards the exact binding momentum $p_{\rm exact}=im\alpha$, such for $n\geq 5$ the difference with respect to the latter is at the level of a few per mil or less. Reference~\cite{Oller:2022tmo} also compared the unitarized $T_0(p)$ with the exact one and this is shown in Fig.~\ref{fig.220711.1}, where the exact result for $|T_0(p)|$ is the (black) solid line, and approximations for $n$ up to 7 are shown.  We see that as $p$ decreases compared to $m\alpha$ more orders are required to reproduce the exact result. This is because the Coulomb scattering becomes trivial for $p\to\infty$, while for $|p|\lesssim 0.6 m\alpha$ the exact $V_0(p)$ has poles in the complex $p$-plane, and its perturbative calculation loses its meaning. 

     With the perturbatively calculated $F_0^{(1)}$ and $F_0^{(2)}$, in terms of a finite but vanishing photon mass, one can apply {\it and test} for Coulomb scattering the method advocated in Ref.~\cite{Dobado:2022} for unitarizing graviton-graviton scattering. In terms of those perturbative PWAs the unitarized formula for $T_0(p)$ reads
     \begin{align}
  \label{221017.6}
T_0(p)&=\frac{F_0^{(1)}(p)^2}{F_0^{(1)}(p)-F_0^{(2)}(p)}~,
     \end{align}
     as an application of the Inverse Amplitude Method \cite{Truong:1988zp,Dobado:1989qm,Oller:1998hw,Oller:1997ng}. However, when taking $\mu\to 0$ the previous formula collapses to
     \begin{align}
       \label{221017.1}
       \lim_{\mu\to 0}T_0(p)=\frac{i2\pi}{mp}~,
     \end{align}
     without any dynamical content, in contradiction with the (black) solid line in Fig.~\ref{fig.220711.1} or with the existence of the ground state of the Hydrogen atom. Therefore, the method of Ref.~\cite{Dobado:2022} fails to reproduce Coulomb scattering and, in general, is not suitable to unitarize infinite range interactions. 

     \section*{Acknowledgements}

I would like to thank Jorge Mart\'{i}n Camalich and Diego Blas for their collaboration in  Refs.~\cite{Blas:2020och,Blas:2020dyg}. 
This work has been supported in part by the MICINN
AEI (Spain) Grant No. PID2019–106080GB-C22/AEI/10.13039/501100011033. 

\bibliography{ref}

\end{document}